\documentclass[aps,numerical,graphicx,showpacs,reprint,prl]{revtex4-1}
\usepackage{amsmath}
\usepackage{graphicx}

\begin{document}

\def\Journal#1#2#3#4{{#1 }{\bf #2, }{ #3 }{ (#4)}}

\def\BiJ{ Biophys. J.}
\def\Bios{ Biosensors and Bioelectronics}
\def\LNC{ Lett. Nuovo Cimento}
\def\JCP{ J. Chem. Phys.}
\def\JAP{ J. Appl. Phys.}
\def\JMB{ J. Mol. Biol.}
\def\JPC{ J. Phys: Condens. Matter}
\def\CMP{ Comm. Math. Phys.}
\def\LMP{ Lett. Math. Phys.}
\def\NLE{{ Nature Lett.}}
\def\NPB{{ Nucl. Phys.} B}
\def\PLA{{ Phys. Lett.}  A}
\def\PLB{{ Phys. Lett.}  B}
\def\PNAS{Proc. Natl. Am. Soc.}
\def\PRL{ Phys. Rev. Lett.}
\def\PRA{{ Phys. Rev.} A}
\def\PRE{{ Phys. Rev.} E}
\def\PRB{{ Phys. Rev.} B}
\def\PD{{ Physica} D}
\def\ZPC{{ Z. Phys.} C}
\def\RMP{ Rev. Mod. Phys.}
\def\EPJD{{ Eur. Phys. J.} D}
\def\SAB{ Sens. Act. B}
\title{
Duality and reciprocity of fluctuation-dissipation relations in conductors}
\author{Lino Reggiani}
\email{lino.reggiani@unisalento.it}
\affiliation{Dipartimento di Matematica e Fisica, ``Ennio de Giorgi'',
Universit\`a del Salento, via Monteroni, I-73100 Lecce, Italy}
\affiliation{CNISM,  Via della Vasca Navale, 84 - 00146 Roma, Italy}

\author{Eleonora Alfinito}
\affiliation{Dipartimento di Ingegneria dell' Innovazione, Universit\`a del
Salento, via Monteroni, I-73100 Lecce, Italy}
\affiliation{CNISM,  Via della
Vasca Navale, 84 - 00146 Roma, Italy}

\author{Tilmann Kuhn}
\affiliation{Institut f\"ur Festk\"orpertheorie, Universit\"at M\"unster,
Wilhelm-Klemm-Str.~10, 48149 M\"unster, Germany}

\date{\today}
\begin{abstract}
By analogy with linear-response we formulate the duality and reciprocity
properties of current and voltage fluctuations expressed by Nyquist relations
including the intrinsic bandwidths of the respective fluctuations. For this
purpose we individuate total-number and drift-velocity  fluctuations of
carriers inside a conductor as the microscopic sources of noise. The spectral
densities at low frequency of the current and voltage fluctuations and the
respective conductance and resistance are related in a mutual exclusive way
to the corresponding noise-source. The macroscopic variance of current and
voltage fluctuations are found to display a dual property via a plasma
conductance that admits a reciprocal plasma resistance. Analogously, the
microscopic noise-sources are found to obey a dual property and a reciprocity
relation. The formulation is carried out in the frame of the grand canonical
(for current noise) and canonical (for voltage noise) ensembles and results
are derived which are valid for classical as well as for degenerate
statistics including fractional exclusion statistics. The unifying theory so
developed sheds new light on the microscopic interpretation of dissipation
and fluctuation phenomena in conductors. In particular it is proven that, as
a consequence of the Pauli principle, for Fermions non-vanishing
single-carrier velocity fluctuations at zero temperature are responsible for
diffusion but not for current noise, which vanishes in this limit.
\end{abstract}

\maketitle 

{\it Introduction} -
The dual property in electrical transport in the linear-response regime
asserts that perturbation (applied voltage $V$ or imposed current $I$) and
response (measured current or voltage drop) can be interchanged with the
associated kinetic coefficients (conductance $G$ or resistance $R$,
respectively), being reciprocally interrelated. According to Ohm's law, for a
homogeneous conductor the dual property gives
\begin{equation}
I = GV \quad \mbox{and} \quad V = R I  ,
\label{IGV}
\end{equation}
with the reciprocity relation given by
\begin{equation}
GR = 1 \ .
\label{GR}
\end{equation}
The most used applications of the above properties are Th\'evenin's and
Norton's theorems that in electrotechnics are two equally valid methods of
reducing a complex linear-network down to something simpler to analyze
\cite{johnson03}.

The aim of this letter is to formulate an analogous dual property and
reciprocity relation for electrical fluctuations at thermal equilibrium. Here
the perturbation is the microscopic source of spontaneous fluctuations inside
a conductor (taken as the physical system), and the response is the variance
of the macroscopic response (i.e., the variance of current or voltage
fluctuations) measured in the outside circuit. The individuation of the noise
sources at a kinetic level, and thus beyond the simple temperature model, is
a major issue in statistical physics that received only partial, and
sometimes controversial, answers even in the basic literature
\cite{johnson28,nyquist28,callen51,kubo66,klimontovich87}. Ultimately, a
unifying theory that applies generally to classical and degenerate
statistics, including explicitly Fermi-Dirac and Bose-Einstein distribution
functions, is not available to our knowledge. Here, all these issues will be
addressed and formally solved in the framework of the basic laws of
statistical mechanics.

For the analysis of current or voltage fluctuations on a kinetic level a
correct system definition becomes of prime importance. On the one hand, the
microscopic model for carrier transport implies a well-defined equivalent
circuit. On the other hand, the measurement of current or voltage
fluctuations in the outside circuit is reflected in the boundary conditions
for the microscopic modeling, which determine the choice of the appropriate
statistical ensemble. Current noise is measured in the outside short-circuit,
which implies an open system where carriers may enter or leave the sample,
thus referring to the grand canonical ensemble (GCE). Voltage noise is
measured in the outside open circuit when the carrier number in the sample is
fixed, thus referring to the canonical ensemble (CE). While it is well-known
that in the thermodynamic limit different statistical ensembles become
equivalent \cite{landsberg54}, this does not hold anymore in the case of
fluctuations, when a finite system size has to be considered. Nevertheless we
will show that the dual property provides a direct link between the noise
sources in the GCE and the CE.

{\it Theory} -
Electrical fluctuations of a conductor in the limit of low frequency (i.e.,
$\omega \rightarrow 0$) are described by the Nyquist relations
\cite{nyquist28}
\begin{eqnarray}
S_I(\omega=0)  &=& 4 \frac{ \overline{I^2}}{\Delta f_I} =4 K_BT G \ ,
\label{SI}  \\
S_V (\omega=0) &=& 4 \frac{ \overline{ V^2}}{\Delta f_V} = 4 K_BT R \ ,
\label{SV}
\end{eqnarray}
where $S_I$ and $S_V$ are the spectral densities of instantaneous current and
voltage fluctuations, respectively, $\overline{I^2}$ and $\overline{V^2}$ the
variances of the corresponding current and voltage fluctuations, (we recall
that being at thermal equilibrium their average values are identically zero),
$\Delta f_I$ and $\Delta f_V$ the corresponding intrinsic bandwidths
determined by the decay of the corresponding correlation functions, $K_B$ is
Boltzmann's constant and $T$ the absolute temperature. Here and henceforth,
the bar over physical quantities denotes the ensemble average. The dual
property we are interested in refers to the above Nyquist relations, also
called \textit{fluctuation-dissipation theorems} (FDTs) \cite{kubo66}.

\begin{figure}
\centering
\includegraphics[width=\columnwidth]{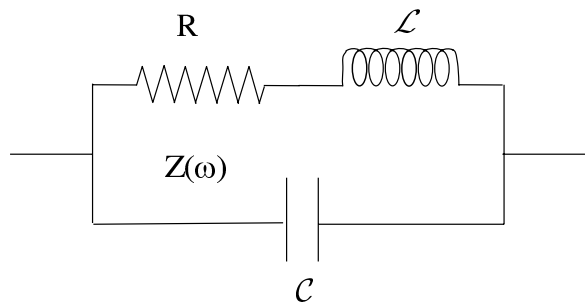}
\caption{Equivalent circuit with impedance $Z(\omega)$ of a real homogenous conductor
of resistance $R$. The capacitance $\cal {C}$ and inductance ${\cal L}$  account
for the presence of the contacts and for the inertia of carriers, respectively.
}
\label{fig1}
\end{figure}

To describe the real conductor on a macroscopic level the ideal resistance
$R$ has to be complemented by a kinetic inductance ${\cal L}$ associated with
the inertia of the carriers and a capacitance ${\cal C}$ associated with the
contacts resulting in the equivalent circuit shown in Fig.~\ref{fig1}. Note
that current and voltage fluctuations are measured under different operation
conditions: Voltage noise is measured at the contacts in the open circuit
(i.e., for $I(t)=0$) while current noise is measured in the outside short
circuit (i.e., for $V(t)=0$).

While the open circuit, being characterized by a closed system with no
particle exchange, is a well-defined concept, the short circuit deserves some
more comments. In order to be an ideal short circuit, the resistance of the
external circuit has to be negligible compared to the resistance of the
conductor under consideration. The resistance of a material is inversely
proportional to the momentum relaxation time and the number of carriers [see,
e.g., Eq.~(\ref{RLC})]. Since the momentum relaxation time typically cannot
be varied to a large extent (unless extremely low temperatures are
considered), the ideal short circuit requires a very large number of
carriers, such that it indeed can be treated as a reservoir in the sense of
the GCE.

On a microscopic level current and voltage fluctuations are generated by the
stochastic motion of the carriers in the conductor. For the reason of clarity
of the model we assume a homogeneous conductor with length $L$ in
$x$-direction and cross-section $A$. The link between microscopic and
macroscopic picture is provided by the Ramo-Shockley theorem
\cite{shockley38,ramo39}, which for our geometry reads
\begin{equation}
\frac{d}{dt} V(t) = \frac{L}{\varepsilon_0 \varepsilon_r A} \left[
\frac{N(t) q}{L} v_d(t) - I(t) \right],
\label{RamoShockley}
\end{equation}
where $\epsilon_0$  and $\epsilon_r$ are the vacuum and the relative
dielectric constant of the host lattice material, respectively, and
\begin{equation}
v_d(t)=\frac{1}{N(t)} \sum_i v_{i,x}(t)=\frac{1}{N(t)} \sum_{\bf k} v_{{\bf k},x}
n_{\bf k}(t)
\label{vd}
\end{equation}
is the instantaneous drift velocity in $x$-direction of the $N(t)$ carriers
with charge $q$ and effective mass $m$ moving with velocities
$\mathbf{v}_i(t)$ in the conductor, the second form being written in terms of
the fluctuating occupation number of carriers $n_{\bf k}(t)$ in the state
${\bf k}$ with the velocity component $v_{{\bf k},x}$ in this state, which
explicitly accounts for the indistinguishability of the carriers. Extensions
of the theorem to more general boundary conditions and quantum mechanical
currents can be found in Refs.~\cite{pellegrini86,pellegrini93}. We notice
that under voltage noise operation (CE) $N(t) = N$ with $N$ the fixed number
of carriers inside the conductor and, for average quantities, the usual
assumption $N=\overline{N}$ is well justified \cite{landsberg54}. This
microscopic model indeed leads to the equivalent circuit shown in
Fig.~\ref{fig1} and, using kinetic theory and a Drude model for the carriers,
its lumped elements are related to the microscopic properties according to
\begin{equation}
R = \frac{L^2 m }{q^2 N \tau}, \quad
{\cal L} = \frac{L^2 m }{q^2 N}, \quad
{\cal C} = \frac{A}{L} \ \epsilon_0 \epsilon_r \ ,
\label{RLC}
\end{equation}
with $\tau$ being the momentum relaxation time. Note that on the microscopic
level the different operation conditions are reflected in different boundary
conditions for the carriers. Voltage noise operation refers to a closed
system with fixed carrier number, while current noise operation refers to an
open system, where carriers enter or leave the system through the contacts.

Within a correlation function scheme the intrinsic bandwidths are directly
related to the decay of the correlation functions associated with the current
and voltage fluctuations. In the case of current noise, $\Delta f_I =1/ \tau
= R / {\cal L}$ is determined by the momentum relaxation time $\tau$ or,
equivalently, by the $R{\cal L}$ time constant of the equivalent circuit.
Analogously, the intrinsic bandwidth of voltage fluctuations $\Delta f_V = 1/
\tau_d = 1/(R{\cal C})$ is determined by the dielectric relaxation-time
$\tau_d=\tau_p^2/ \tau$ or, equivalently, by the $R{\cal C}$ time constant of
the circuit. Here, $\tau_p$ denotes the plasma time (i.e., the inverse of the
plasma frequency $\omega_p$)
\begin{equation}
\tau_p=\omega_p^{-1} =\sqrt{\frac{\epsilon_0 \epsilon_r m A L}{Nq^2}} = \sqrt{{\cal LC}}
\label{taup}
\end{equation}
being related to the ${\cal LC}$ time constant of the circuit. We remark that
also outside the static limit (i.e., for $\omega \ne 0$) the equivalent
circuit in Fig.~\ref{fig1} gives the impedance (or the admittance) whose real
parts reproduce the frequency dependence of the current and voltage spectra
in the classical limit $K_BT \gg \hbar \omega$, with $\hbar$ being the
reduced Planck constant. These spectra are associated with the microscopic
time scales of the corresponding correlation functions, as was validated by
Monte Carlo simulations \cite{reggiani13}. Furthermore, the equivalent
circuit consistently recovers the standard relations
\begin{equation}
\overline{V^2} = \frac{K_BT}{{\cal C}} \quad \mbox{and} \quad
\overline{I^2} = \frac{K_BT}{{\cal L}}.
\label{variances}
\end{equation}
which in this form are valid for any type of inductance and capacitance in a
circuit as given in Fig.~\ref{fig1}. As such, this circuit is of most
physical importance and should replace alternative equivalent circuits (like,
e.g., simple $R \cal C$ parallel and $R{\cal LC}$ serial circuits) which are
also sometimes used in the literature.

Making use of statistics, the temperature can be associated with the
microscopic quantities defining two noise sources that are present in the
conductor, one for each operation condition. These sources should be taken as
mutually exclusive for each of the two boundary conditions. Accordingly,
constant-voltage (i.e., current noise) operation is associated with a GCE
reflecting the open system where \cite{lax60}
\begin{equation}
\overline{\delta N^2} = K_B T \frac{\partial \overline{N}}  {\partial \mu_0\ } .
\label{DeltaN}
\end{equation}
Here, $\overline{\delta N^2}$ is the variance of the instantaneous number of
carriers inside the sample, and $\mu_0$
the chemical potential. Constant-current (i.e., voltage noise) operation, on
the other hand, is associated with a CE with fixed carrier number ($N
=\overline{N}$) and
\begin{equation}
\overline{\delta v_d^2} =
\frac{1}{N^2} \sum_{\bf{k}} v_{{\bf k},x}^2 \  \overline{\delta f^2(\varepsilon_{\bf{k}})} ,
\label{Deltavddef}
\end{equation}
where $\overline{\delta v_d^2}$ is the variance of the fluctuations of the
instantaneous carrier drift-velocity averaged over the sample, $v_{{\bf
k},x}=\hbar k_x / m$ is the $x$-component of its velocity,
$\varepsilon_{\bf{k}}$ is the corresponding energy, $\overline{ \delta
f^2(\varepsilon_{\bf{k}})}=\overline{ n_{\bf k}^2} - \overline{n}_{\bf k}^2$
is the variance of the occupation number and $f(\varepsilon_{\bf{k}}) =
\overline{n}_{\bf k}$ is the equilibrium distribution-function normalized to
carrier number which, according to statistics, satisfies the property
\begin{equation}
\overline{\delta f^2(\varepsilon_{\bf{k}})} = -K_B T \frac{\partial f(\varepsilon_{\bf{k}})}
{\partial \varepsilon_{\bf{k}}} \ ,
\end{equation}
Using the symmetry of the problem, $v_{{\bf k},x}^2$ can be replaced by
$(2\varepsilon_{\bf{k}}/md)$, where $d$ denotes the dimension of the system.
With the density-of-states of a $d$-dimensional carrier gas satisfying ${\cal
D}(\varepsilon) \propto \varepsilon^{(d/2)-1}$, Eq.~(\ref{Deltavddef}) can be
evaluated to
\begin{equation}
\overline{\delta v_d^2} = \frac{K_B T}  { m \ N} \ ,
\label{Deltavd}
\end{equation}
independent of the dimension $d$. We remark that since these noise sources
are directly obtained from the properties of the statistical ensembles, they
hold for carriers obeying any type of statistics, in particular for
Fermi-Dirac (FD) and for Bose-Einstein (BE) statistics, but also for carriers
obeying a fractional exclusion statistics (FES) \cite{halperin84,wu95}. In
all cases Maxwell-Boltzmann (MB) statistics is implicitly recovered in the
limit $f \ll 1 $.

By using Eq.~(\ref{DeltaN}) to replace the temperature in Eq.~(\ref{SI}) the
current fluctuations can be expressed as
\begin{equation}
\overline{I^2} = \frac{G}{\tau} \, \overline{\delta N^2} \, \frac{\partial \mu_0}
{\partial \overline{N}} \ .
\label{I2}
\end{equation}
By taking $G$ from the generalized Einstein relation \cite{gurevich79}
\begin{equation}
G = \left(\frac{q}{L}\right)^2 D   \frac{\partial \overline{ N}} { \partial \mu_0} \ ,
\label{GD}
\end{equation}
where
\begin{equation}
D=\overline{v^{2'}_x} \tau
\label{diffcoeff}
\end{equation}
is the longitudinal-diffusion coefficient \cite{gurevich79} with the
differential (with respect to carrier number) quadratic velocity component
along the $x$-direction given by
\begin{eqnarray}
\overline{v^{2'}_x} &=& \sum_{\bf{k}} v^{2}_{{\bf k},x}  \,
\frac{\partial f(\varepsilon_{\bf{k}})} {\partial \overline{N}} \nonumber \\
&=& \sum_{\bf{k}} v^{2}_{{\bf k},x}  \,
\frac{\partial f(\varepsilon_{\bf{k}})} {\partial \mu_0}
\frac {\partial \mu_0} {\partial \overline{ N}} =
\frac{K_B T}{m} \frac {\overline {N}}  {\overline {\delta N^2}} \ ,
\label{vx2p}
\end{eqnarray}
where the last equality, obtained in the same way as
Eq.~(\ref{Deltavd}), follows from Eq.~(\ref{Deltavddef}) and additionally
using Eq.~(\ref{DeltaN}). Notice the explicit appearance of the Fano factor,
$\overline {\delta N^2} / \overline {N}$, to account for the effective
interaction among carriers due to the symmetry properties of their wave
functions and thus for the correct statistics.

Then, Eq.~(\ref{I2}) takes the equivalent forms
\begin{equation}
\overline{ I^2} = \frac{K_B T}{{\cal L}}
= \frac{q^2 \overline{v^{2'}_x} \ \delta\overline{ N^2}} {L^2 }
= \frac{ q^2  \delta\overline{ N^2}}{\tau_N^2}
\label{I2tauN}
\end{equation}
with $\tau_N = \sqrt{L^2 / \, \overline{v^{2'}_x}}$ being an effective
transport-time through the sample \cite{greiner00} determining the conversion
of total-number fluctuations of carriers inside the sample into total-current
fluctuations measured in the external short-circuit.

For the variance of voltage fluctuations, substitution of Eq.~(\ref{Deltavd})
into Eq.~(\ref{SV}) gives the equivalent expressions
\begin{equation}
\overline{V^2}  =
\frac{K_B T} {{\cal C}}
=\frac{m N L \overline{\delta v_d^2}} {A \epsilon_0 \epsilon_r}
=\frac{m^2 L^2 \overline{\delta v_d^2}} {q^2 \tau_p^2}
\label{V2}
\end{equation}
with the plasma time $\tau_p$ given in Eq.~(\ref{taup}). Introducing the
variance of electric-field fluctuations averaged over the sample length
$\overline{\delta E^2} = \overline{\delta V^2}/L^2$ and the plasma
carrier-mobility $\mu_p=q\tau_p / m$, Eq.~(\ref{V2}) can be written in terms
of a generalized Ohm's law
\begin{equation}
\overline{\delta v_d^2} = \mu^2_p \, \overline{\delta E^2} \ ,
\label{DeltaE2}
\end{equation}
describing the conversion of carrier drift-velocity fluctuations inside the
sample into electric field  (or voltage) fluctuations at the terminals of the
open circuit.

By using  Eqs.~(\ref{SI}) and (\ref{DeltaN}) the macroscopic conductance is
associated with carrier total-number fluctuations by:
\begin{equation}
G = \frac{q^2 \overline{v^{2'}_x} \tau} {L^2K_BT} \  \overline{ \delta N^2}.
\label{GN2}
\end{equation}
Analogously, using Eqs.~(\ref{SV}) and (\ref{Deltavd})  the macroscopic
resistance is associated with drift-velocity fluctuations by:
\begin{equation}
R = \frac{L^2 m^2} {q^2 \tau K_BT } \overline{\delta v_d^2} .
\label{Rvd2}
\end{equation}
From a microscopic point of view, Eqs.~(\ref{GR}), (\ref{GN2}) and
(\ref{Rvd2}) imply that the noise sources satisfy the relations
\begin{equation}
GR = \frac{\overline{v^{2'}_x} \overline{\delta N^2} m^2 \overline{\delta v_d^2}   } {(K_BT)^2  }
= \frac{\overline { v^{2'}_x}} {\overline{\delta v_d^2}} \
\frac{\overline{\delta N^2}}{N^2} = 1
\label{GR2}
\end{equation}
and thus
\begin{equation}
\overline{ \delta N^2}  \ \overline { v^{2'}_x}  =
{ N}^2  \ \overline{ \delta v_d^2} \
= \frac{NK_BT}{m}  ß ,
\label{N2v2}
\end{equation}
where again Eq.~(\ref{Deltavd}) has been used to eliminate the temperature
in Eq.~(\ref{GR2}).

Equations (\ref{GR2})  and (\ref{N2v2}) express the reciprocity and duality
properties of the microscopic noise-sources associated with the
fluctuation-dissipation relations. In other words, at thermodynamic
equilibrium carrier total-number fluctuations inside a conductor under
constant-voltage conditions are inter-related to carrier  drift-velocity
fluctuations under constant-current conditions.

The dual property of the macroscopic FDTs is obtained from
Eqs.~(\ref{I2tauN}), (\ref{V2}) and (\ref{N2v2}) as
\begin{equation}
\overline{I^2} = G^2_p \ \overline{V^2} \quad \mbox{and} \quad
\overline{V^2} = R^2_p \ \overline{I^2}
\label{I2V2}
\end{equation}
with a plasma conductance $G_p=(qN\mu_p)/L^2$ and a plasma resistance $R_p$
satisfying the reciprocity relation
\begin{equation}
G_p R_p=1 \ .
\label{GpRp}
\end{equation}
By satisfying the relations (\ref{SI}) and (\ref{SV}), the expressions
(\ref{I2V2}) and (\ref{GpRp}) justify the identification of the intrinsic
bandwidths here assumed.

Equations (\ref{I2V2}) and (\ref{GpRp}) express the duality and reciprocity
properties of fluctuation-dissipation relations and parallels
Eqs.~(\ref{IGV}) and (\ref{GR})  of linear-response relations. Notice, that
all the above expressions hold for any type of statistics (in the case of
Bosons for temperatures above the critical temperature for Bose-Einstein
condensation \cite{davies68}), thus complementing the standard FDT in the
limit of low frequencies. From statistics, the two boundary conditions are
associated with a GCE and a CE, respectively, and Eq. (\ref{N2v2}) shows the
interesting results that both statistics provide the same result even outside
the thermodynamic limit conditions \cite{landsberg54}.

{\it Conclusions} -
We have formulated the dual property and the reciprocity relation of the FDTs
in the limit of low frequency by expressing conductance and resistance in
terms of the proper microscopic noise-source, what we call the
dissipation-fluctuation relations. Analogously, by accounting for the
intrinsic bandwidth, the variances of current and voltage fluctuations are
related to the same noise-sources, what are usually called the fluctuation
dissipation relations. All these relations are given in a form that is
independent of the type of distribution functions, thus including MB, FD and
BE statistics (the latter at temperatures above Bose-Einstein condensation)
as well as FES. From a physical point of view, the temperature entering the
Nyquist relations (\ref{SI}) and (\ref{SV}) is here expressed in kinetic
terms and associated with the variance of instantaneous fluctuations (i) of
the total number of carriers inside the sample, or (ii) of the carrier
drift-velocity inside the sample. Calculations are carried out in the
framework of the GCE and the CE and the result summarized in Eq.~(\ref{N2v2})
provides an interesting example where both ensembles  give the same result
even outside the thermodynamic limit.

While thermal noise and shot noise are typically described as different noise
phenomena, one originating from the thermal agitation of the carriers
\cite{johnson28,nyquist28} and the other from the discreteness of the charge
\cite{schottky18,blanter00,beenakker03}, our results clearly indicate the
close relationship between both of them. Indeed, by expressing current
fluctuations in terms of the ratio between the variance of carrier number
fluctuations and the effective transport time $\tau_N$ we show that the
source of shot-noise is already present at thermal equilibrium
\cite{gomila04}.

The present formulation shows that for Fermions the vanishing of the
low-frequency current spectral-density at $T=0$ is associated with the
vanishing of the variance of carrier total-number fluctuations, i.e., with
the instantaneous correlation between appearance and disappearance of an
elemental carrier number fluctuation inside the sample. This is essential
because for Fermions the value of the diffusion coefficient
[Eq.~(\ref{diffcoeff})] does not vanish at $T=0$, and thus the notion of
diffusion as synonymous of noise, fails completely. In contrast, for the
classical case the absence of motion at $T=0$, i.e., $D=0$ and $\tau_N
\rightarrow \infty$, is definitely responsible for the vanishing of current
noise.

For the case of voltage fluctuations, the vanishing of thermal noise at
$T=0$ is associated with the tendency of the drift-velocity fluctuations to
approach zero, which is a property independent of the kind of statistics.

As a final remark, we notice that the equivalent circuit here introduced
reproduces the correct frequency dependence of both current and voltage
spectral densities, the latter one including the plasmonic contribution in
the case $\tau_d \ll \tau$ \cite{reggiani13}. We want to stress that the time
scales of the fluctuating macroscopic variables are in general not related to
those of the respective noise sources. Instead, they should be treated in the
framework of the time or frequency dependence of the corresponding
correlation functions or spectral densities.

\end{document}